\documentclass[12pt]{article}
\usepackage{latexsym}            % om mooie symbolen te maken
\usepackage{epsf}
\usepackage[dvips]{graphicx}
\input epsf.tex

\def\be{\begin{equation}}
\def\ee{\end{equation}}
\def\bea{\begin{eqnarray}}
\def\eea{\end{eqnarray}}
\def\ba{\begin{array}}
\def\ea{\end{array}}

\def\part{\partial}

\def\makeatletter{\catcode`\@=11}% 11:letter
\makeatletter
\def\mathbox#1{\hbox{$\m@th#1$}}%
\def\math@ccstyles#1#2#3#4#5#6#7{{\leavevmode
      \setbox0\mathbox{#6#7}%
      \setbox2\mathbox{#4#5}%
      \dimen@ #3%
      \baselineskip\z@\lineskiplimit#1\lineskip\z@
      \vbox{\ialign{##\crcr
             \hfil \kern #2\box2 \hfil\crcr
             \noalign{\kern\dimen@}%
             \hfil\box0\hfil\crcr}}}}
\def\mathaccstyles{\math@ccstyles\maxdimen}
\def\maththroughstyles{\math@ccstyles{-\maxdimen}}
\def\unity%
 {\maththroughstyles{.45\ht0}\z@\displaystyle
 {\mathchar"006C}\displaystyle 1}
\begin{document}
%\pagestyle{empty}

%titel
\rightline{December 2007} \vspace{1truecm}

%%%%%%%%%%%
\centerline{\Large \bf On Uniqueness of supersymmetric Black holes in AdS(5)} \vspace{1truecm}

\centerline{
    {\bf Pedro J. Silva${}$}\footnote{E-mail address:
                                  {\tt psilva@ifae.es}}}

\vspace{.4truecm} \centerline{{Institut de Ci\`encies de l'Espai (IEEC-CSIC) and
Institut de F\'{\i}sica d'Altes Energies (IFAE),}} \centerline{{\it UAB, E-08193 Bellaterra (Barcelona), Spain.}}
\vspace{2truecm}

%%%%%%%%%%%%%%%%
\centerline{\bf ABSTRACT}
\vspace{.5truecm}

\noindent We study the possibility of having Black hole of spherical and ring horizon topology with five independent charges in the $U(1)^3$-model of 5D gauge supergravity. To study these possibilities we consider not only the known result obtained by local supersymmetry analysis but include the input coming from non-local properties of the solutions, like the attractor mechanism, the entropy function of Sen, the Euclidean formulation and general properties of the uplift to ten dimension. For the spherical case, we found that there is no room for more general Black holes than the ones already describe in hep-th/0601156. On the other hand, if a solution of ring horizon topology exists, we conclude that it must be labeled by three independent parameters only, since it has to satisfy two independent constraints that we explicitly find in terms of its chemical potentials. At the end of the article, based on all the local and non-local information, we put forward a conjecture on the constraints that characterize general Black holes dual to ${\cal N}=4$ SYM.

%%%%%%%%%%%%%%%%%%%%%%%%%%%%%%%%%%%%%%%%%%%%%%%%%%%%%%%%%
%\newpage

\section{Introduction}

\noindent String theory has provided the first microscopic derivation
of the entropy for some extremal Black holes (Bh) in diverse dimensions (see for example \cite{Strominger:1996sh,Emparan}). All the calculations are based on the identification of a given set of microscopic states forming an ensemble (label by the Bh charges) with the Bh geometry. In four dimensions we have Bh uniqueness theorems which guaranty that to a given set of asymptotic charges there corresponds only one Bh solution. This is in general not true for higher dimension where such theorems no longer exist. In particular in five dimensions we have Black rings (Br) and spherical Bh solutions as an explicit example of non-uniqueness. Considering the above, it seems important to understand better what kind of Bh exist in string theory compactifications, if we want to understand to which microscopic ensembles these Bh are related.  Unfortunately, this is a very difficult problem to solve in general that nevertheless can be significantly simplified if we include supersymmetry as an input.

In fact, supersymmetric Bh in minimal five dimensional
supergravity and its extension to the $U(1)^n$ vector multiples
were classified in \cite{Reall:2002bh,Gutowski:2004bj}. Here, local analysis of supersymmetry was used to classify all the near horizon geometries (NHG) to
later match them with asymptotic data that defines the Bh
solutions. The main result in the minimal cases is the uniqueness of the corresponding supersymmetric BMPV Bh \cite{Breckenridge:1996is} and the supersymmetric Br solutions of \cite{Elvang:2004rt}.

In this article, we consider the very important framework of ten
dimensional type IIB supergravity with $AdS_5\otimes S^5$
asymptotic conditions, related by duality to $U(N)$
${\cal N}=4$ super Yang-Mills in four dimensions. This theory can
be consistently truncated and compactified on the $S^5$ to give
$SO(6)$ gauge supergravity in five dimensions, that furthermore
can be truncated to the $U(1)^3$-model that also admits a
final truncation to minimal gauge supergravity. These different
truncations tell us that we can have Bh solutions at four
different levels.
\begin{enumerate}
\item Bh in 10D type IIB supergravity.
\item Bh in $SO(6)$ 5D gauge supergravity.
\item Bh in $U(1)^3$ 5D gauge supergravity.
\item Bh in minimal 5D gauge supergravity.
\end{enumerate}

All supersymmetric Bh should be describe in the dual CFT by
supersymmetric ensembles of states label by the quantum charges
corresponding to the $SO(4)$ angular momenta $J_i$, $i=(1,2)$ and
the SO(6) R-charges $Q_I$, $I=(1,2,3)$. At present, there only
exist explicit supersymmetric Bh solutions in the last two
truncations (3) and (4) \cite{sbh1,sbh2,sbh3}, but is not clear if these
solutions are the most general Bh or just a particular family. The
key point is that these solutions come with a constraint among
the conserved charges that is not apparent in the dual theory
\cite{index}.

Regarding the classification of Bh solutions in this framework,
only truncation (3) and (4) have been studied so far \cite{fhp,NH1,NH2,Ast,NH3}, where
unfortunately local analysis of supersymmetry was too complicated to
be fully solved, in both cases. More specifically, only the NHG
with two $U(1)$ symmetries are classified and the necessary
connection with the corresponding asymptotic region was not
achieved, leaving the classification unsettle.

In this work we propose a complementary approach to help on the
classification  of supersymmetric Bh solutions, bypassing the
problem of the asymptotic form of the metric. We base our
construction on global properties of supersymmetric Bh like the
attractor mechanism \cite{attractor}, its Euclidean formulation,
thermodynamics \cite{yo1,yo2} and general properties of the uplifting
of gauge supergravities to 10D \cite{g}. We explicitly consider only the last two truncations, but these new ideas should help also in the other more general cases and in fact we discuss this possibility and some of its implication in the last section.

Our main results are that in the $U(1)^3$-model, there are no other BPS Bh with $S^3$ horizon than the ones we already know, and that if there is a BPS Br in $AdS$ it is highly constraint with only 3 independent parameters. Even more, for these Br, there seems to be an obstruction to take the un-gauge limit (sending the $AdS$ radius to infinity), as the chemical potentials conjugated to the dipole momenta become unbounded. More generally, we found that all BPS Bh in the $U(1)^3$-model present the same single constraint among its chemical potentials. Finally, and  at a more speculative level, we propose as a conjecture, that all BPS Bh dual to ${\cal N}=4$ SYM show this same constraint, that is the same relation that appears in the definition of the SYM Index of \cite{index}.

The plan of this work is the following, we summarize the main results of local supersymmetric analysis in section \ref{localsusy}. Then in section \ref{attractors}, we consider the information that can be extracted from the attractor mechanism and the entropy function of Sen. In section \ref{uplift}, we use the extension of the usual thermodynamics relations, the Euclidean formulation and the uplift to 10D of 5D solutions, to derive new more general global constraints of the Bh geometries. Finally, in section \ref{end}, we summarize our result, discuss on future directions and formulate the above conjecture on BPS Bh constraints.

\section{Supersymmetry \& local analysis}
\label{localsusy}

Analysis of local supersymmetry based on killing spinors is a
powerful technique to classify supersymmetric solutions (see for example \cite{clasification}). This method was adapted in \cite{Reall:2002bh} to study the NHG of putative Bh solutions for un-gauge supergravity. Later in \cite{NH1,NH3}, this method was applied first to minimal gauge supergravity and then to its
$U(1)^n$ vector multiplet extension. Although the full classification of Bh geometries was not obtained, due to the complicated structure of the theory, the NHG were classified up to one rotational isometry. The main result in the $U(1)^3$-model is that there exist three types of regular NHG corresponding to compact horizons;
\begin{itemize}
\item $AdS_2\otimes S^3_{squashed}$ ,
\item $AdS_2\otimes S^2\otimes S^1$,
\item $AdS_2\otimes T^3$.
\end{itemize}
The first case corresponds to BH with topological
spherical horizons while the second and third to BH with
topological $S^2\otimes S^1$ and $T^3$ horizons. In the original
derivation it is assumed that the Bh have isometry group $R\times
U(1)\times U(1)$, associated to time flow and two rotational
symmetries\footnote{The existence of two $U(1)$ symmetries is
not necessary on general grounds, but they are present in all
known Bh solutions of gauge and un-gage 5D supergravity. On the
other hand in \cite{Gutowski:2007ai} all 1/2 BPS solutions where classified
finding solutions with only one $U(1)$. Therefore the status of
these second symmetry is unclear at this point.}. All these NHG
are 1/2 BPS and depend at most on {\it four} parameters only.

Base on the above partial classification, Bh in this gauge supergravity of the $U(1)^3$-model could appear in the following different situations,
\begin{enumerate}
\item There are more general Bh with only one $U(1)$ rotational
isometry,
\item There are more general Bh with $U(1)^2$ and more parameters
that disappear in the NHG,
\item The Bh we already know are the most general family of solutions with horizon of $S^3$ topology apart from Br solutions and Bh with horizon of $T^3$ topology.
\end{enumerate}
Regarding the truncation to minimal gauge supergravity, after a
short analysis it is easy to see that the second and third cases
can not be reduced from the $U(1)^3$-model, since they have
constraints incompatible with the truncation. We can safely say that there are no supersymmetric Br solutions in minimal gauge supergravity but
there could be supersymmetric Br in the STU-model. On the other
hand, the case of spherical topology admits reduction to minimal
gauge supergravity, telling us that there can be supersymmetric Bh
in both theories. Summarizing in minimal gauge supergravity we are left with the following possibilities,
\begin{enumerate}
\item There are more general Bh with only one $U(1)$ rotational
isometry,
\item There are more general Bh with topological $S^3$ horizon and more parameters that disappear in the NHG,
\item The Bh we know are the most general family of solutions.
\end{enumerate}
It is important to notice that in \cite{NH3}, it was proved that the NHG of the most general known Bh of \cite{sbh3} coincides with the case of $AdS_2\otimes S^3_{squashed}$ NHG for the $U(1)^3$-model and also to its truncation to minimal gauge supergravity.

The above information summarizes the results and possibilities that are known using supersymmetry and local analysis. We need to complement this results with extra input data if we want to fully classify all Bh solutions. In the rest of this work we consider the constraints imposed by the attractor mechanism \cite{attractor} and the entropy function of Sen \cite{sen}, together with global properties necessary to define the Euclidean regime in the uplifted 10D metric to find new restrictions that help in the Bh classification.

In the following section we study the NHG using the attractor mechanism, and hence we mainly focus on the cases assuming Bh with isometry group $R\times U(1)\times U(1)$ unless we explicitly say the contrary.

\section{The attractor mechanism \& the entropy function}
\label{attractors}

The attractor mechanism was originally discussed for ${\cal N}=2$
Bh in four dimensions, where the values of the scalar fields at
the horizon are given by the values of the Bh conserved charges
and are independent of the asymptotic values of the scalars at
infinity. Importantly, the attractor mechanism has provided a new
way to calculate the Bh entropy. In a series of articles
\cite{sen}, Sen recovered the entropy of  $D$-dimensional BPS Bh using only the near horizon part of the geometry. Basically, in this regime the solution
adopts the form $AdS_2 \otimes \Upsilon^{D-2}$ with $\Upsilon$ a compact manifold of the form $\Pi_{n_i}(\times S^{n_i})$ with $\sum n_i=D-2$, plus some electric and magnetics fields. The entropy $S$ is obtained by introducing a
function $f$ as the integral of the corresponding supergravity
Lagrangian over the $\Upsilon^{D-2}$. More concretely, an entropy
function is defined as $2\pi$ times the Legendre transform of $f$
with respect to the electric fields $e$ and rotational parameters
$\alpha$ \footnote{The analysis of the near horizon geometry has
been extended to stationary Bh that define squashed $AdS_2\otimes
S^{D-2}$ geometries in last article of \cite{sen}.}. Then, an
extremization procedure fixes the on-shell BPS values of the
different fields of the solution and in particular, determines
the BPS value of the entropy,
 \bea S_{bps}=2\pi\left( e {\partial f\over
\partial e}+\alpha{\partial f\over
\partial \alpha} -f \right)_{bps}\,.\label{ae}
\eea
More recently the extension of the entropy functional to gauge
supergravities has been considered in \cite{Morales,Suryanarayana:2007rk,yo2,Astefanesei:2007vh}. Here, the characteristic Chern-Simons term is included into the discussion and again the entropy of the Bh is recovered with the sole information
of the NHG.

\vspace{.5cm}
\noindent{\bf Entropy for BPS Bh in $U(1)^3$-model} \vspace{.5cm}

Let us start with the most general Bh solution possessing isometry group $R\times U(1) \times U(1)$. This solution is labeled by a given set of
parameters $l_I$, that can be related to physical quantities $L_I$
among which we have the five conserved charges
$(J_1,J_2,Q_1,Q_2,Q_3)$ \footnote{Other physical quantities
like possible dipole or higher momenta should not be excluded from the list.}. We are interested in its entropy $S_{Bh}$, that is in principle a function of all this parameters i.e. $S_{Bh}(l_{I})$. Let us focus now in the NHG of this Bh solution (that has to be one of the list presented in previous section). As we said before, all these geometries
depend at most in four parameters that we call $\lambda_i$. Using
the entropy functional method we can always calculate
the entropy $ S_{NHG}$ using just the NHG and hence it depends only on the parameters $\lambda_i$ that we write as $ S_{NHG}(\lambda_i)$. On the other hand, by construction both expressions have to be equal i.e.
\be S_{Bh}(l_I)= S_{NHG}(\lambda_i)\,.\ee
The key point of this argument is to realize that we are looking for Bh solutions that are \textit{more general} than the one we already known. In all the known solutions the entropy $S_{Bh}$ is a function of all five conserved charges, and therefore the assumed more general solution should
also depends on all of the conserved charges plus maybe other physical quantities. But we have established that there are only four independent parameters at most, therefore we concluded that the putative more general Bh is a constraint system where the conserved charges are all related by a single equation on the top of the BPS equation. This is our first new result, that in other words says

\vspace{.5cm} \centerline{\textit{I. There is no BPS Bh with isometry group $R\times U(1) \times U(1)$} and five independent}{\textit{\hspace{.5cm} charges within the $U(1)^3$-model.}}\vspace{.5cm}

\noindent {\bf Conserved charges for BPS Bh} \vspace{.5cm}

Let us focus on the conserved charges that can be calculated in these NHG. In \cite{Suryanarayana:2007rk} it was showed how to compute the conserved charges of a given Bh, using only the NHG. The computation is based on a
Noether procedure of the reduced three dimensional solution along
the two $U(1)$ isometries. We should nevertheless recall that the
calculation of the conserved charge is based on a regular behavior
of the solution as we approach the Bh from infinity, to be more
precise, we need a smooth fibration on the radial direction
between the horizon and the corresponding space-like three-surface
at infinity (see section 3 of \cite{Suryanarayana:2007rk}). This is not a strong
condition for squashed horizons and in fact is satisfied for all
known AdS Bh, but things get more complicated for other
topologies. For example, in flat space Br do not satisfy this
constrain and therefore the NH analysis does not provide the form of the conserved charges \footnote{We thanks the authors of \cite{Ha} to point out that in this work they explained how to calculate the electric charges and angular momenta for supersymmetric Br in flat space, base only on the NHG. The procedure is base on using more than one patch of coordinates to overcome the local infinities of the the gauge potentials. It will be interesting to check is implementation for the AdS case.} . Nevertheless, if we assume the above condition to hold, the resulting
expression for the conserved charges is not too difficult to calculate in the NHG with topology of squashed $S^3$. Then, is easy to verify that the NH charges reproduce exactly the same parametric dependence that the five charges of the known Bh solution. In other words,

\vspace{.5cm} \centerline{\textit{II. All BPS Bh with isometry group $R\times U(1) \times U(1)$} and horizon of
topology $S^3$}{\textit{\hspace{.9cm}have the same constraint
among its conserved
charges.}}\vspace{.5cm}

\noindent Unfortunately, we can not extend the above result to other horizon topologies since we know that the NHG may not provide all the information to compute the Bh asymptotic charges. \vspace{.5cm}

\noindent {\bf Chemical potentials for BPS Bh in AdS} \vspace{.5cm}

On the other hand, we can always characterize the constraints of a given Bh geometry in terms of different set of thermodynamic variables by changing ensembles. In particular, we switch to the Grand canonical ensemble where all charges have been traded for its conjugated potentials. In this case the Euclidean action $I_{Bh}$ looks for example like,
\be I_{Bh}(\phi_I,w_i)=\phi_IQ^I+w_iJ^i-S_{Bh} \label{qsr}\ee
where $(\phi_I,w_i)$ are the chemical potentials conjugated to the
accompanying conserved charges. In \cite{yo1,yo2} it was developed a method to calculate the above quantities for any extremal Bh.

The main point of working in this ensemble, is that the calculation of the conjugated chemical potentials can be carried out entirely using only the NHG, and does not require extra assumptions on the regularity of any fibration on the radial direction. These different chemical potentials turn out to be identify with the different electric fields appearing in the 3D reduction of the 5D solution (see \cite{yo2} for a derivation).

To perform explicitly this reduction along the two $U(1)$ isometries on the different NHG, we follow the notation of \cite{Suryanarayana:2007rk}. In 3D we end up with five $U(1)$ gauge fields $(a^I,B^i)$ where $(I=1,2,3)$ is the R-charge index and $(i=1,2)$ corresponds to the two compactified directions. Explicitly the 5D metric $G_{\mu\nu}$ and gauge fields $A^I_\mu$ are rewritten as,
\bea G_{\mu\nu}=\begin{array}{cc}
                  \left( \begin{array}{cc}
                  g_{MN}+h_{ij}\,b^{\,i}_Mb^{\,j}_N & h_{in}b^{\,i}_M\\
                  h_{im}b^{\,i}_N & h_{mn}
                  \end{array} \right) &\hbox{and}\quad A^I_\mu=
                  (a^I_M+A_ib^{\,i}_M ,A_n)\,,
            \end{array}
\eea where Greek indices are five dimensional and split into
capital Roman indices $(M,N,\ldots)$ corresponding to
$(t,r,\theta)$ and lower Roman indices corresponding to the two
compactified dimensions $(\zeta_1,\zeta_2)$. In particular, the
different electric fields come from the field strength defined as
$H^i=db^i$ and $F^I=da^I$.

\vspace{.5cm}
{\textbf{The $S^3$ case}: Here the physical distinguishing properties of the constraint system is not related to the difference of the two compactified angular directions $(\zeta_1,\zeta_2)$ and therefore it is enough to perform a single reduction along a new angular coordinate defined as $\varphi=\zeta_1+\zeta_2$. Due to the above fact and to simplify the form of the expressions involved, we work only with the NHG with two equal angular momenta. The resulting squashed $S^3$ NHG can now be
recast in terms of three independent parameters $\mu_I$ as follows,
\bea
&&ds^2=v_1\left(-r^2dt^2+{dr^2\over r^2}\right)+v_2\left[\sigma_1^2+
\sigma_3^2+v_3(\sigma_3-\alpha r dt)^2\right]\nonumber \\
&&A^I=-e^Irdt+p^I(\sigma_3-\alpha r dt)\quad,X_I=u_I
\eea
where $\sigma_1=\sin\varphi d\theta-\sin\theta \cos\varphi d\psi$,
$\sigma_2=\cos\varphi d\theta+\sin\theta \sin\varphi d\psi$,
$\sigma_3=d\varphi+\cos\theta d\psi$ and $\theta=(0,\pi)$,
$\varphi=(0,4\pi)$ and $\psi=(0,2\pi)$. The form of the solution
in terms of three independent parameters $\mu$ is,
\bea
&&u_I={\mu_I\over\gamma_3^{1/3}}\,,\quad v_1={\gamma_3^{1/3}\over
4(1+\gamma_1)}\,,\quad v_2={\gamma_3^{1/3}\over 4}\,,\quad
v_3=1+\gamma_1-{\gamma_2^2\over 4\gamma_3}\,,\quad  \nonumber\\
&&\alpha={\gamma_2\over (1+\gamma_1)\sqrt{4\gamma_3(1+\gamma_1)-\gamma_2^2}}\,,
\quad p_I={(\gamma_1-\mu_I)\mu_I^2-\gamma_3\over 4\mu_I^2}\,,\nonumber \\
&&e_I={\gamma_2(\mu_I^3-\gamma_1\mu_I^2)+
[4\gamma_3(1+\gamma_1)-\gamma_2^2]\mu_I + \gamma_2\gamma_3\over
4\mu_I^2(1+\gamma_1)\sqrt{4\gamma_3(1+\gamma_1)-\gamma_2^2}}\,,
\eea
where $\gamma_1=\sum_I\mu_I$, $\gamma_2=\sum_{I<J}\mu_I\mu_J$
and $\gamma_3=\mu_1\mu_2\mu_3$. At this point is easy to read from
the above parametrization, the four different $U(1)$ gauge fields
of the reduced 4D solution,
\bea a^I=-e_I r dt+p_I\sigma_3,\qquad
b=-\alpha r dt+\cos\theta d\psi
\eea
and from these expressions, we extract the final parametrization
of the chemical potentials conjugated to the three R-charges $Q_I$
and the sum of the angular momenta $J_+=(J_1+J_2)$,
\bea
\begin{array}{ccc}
    \left( \begin{array}{cc}
            Q_I \\  J_+
            \end{array}
    \right) &\longleftrightarrow&  \left( \begin{array}{cc}
            \phi_I=e_I \\  2w_+=\alpha
            \end{array}
    \right)\,.
\end{array}
\eea
In terms of these chemical potentials it is not difficult to see that indeed there is a constraint that can beautifully be written as
\be \sum_I \phi_I= 2w_+\,. \label{cc}\ee
This constraint has to be satisfied by the putative general BPS Bh in $AdS$ and as a check, is exactly the same constraint observed in the known BPS AdS Bh solutions of \cite{sbh3}, where we have naturally identified the NH rotational symmetries with the asymptotic rotational symmetries of $AdS$. Therefore we arrive to the conclusion that

\vspace{.5cm} \centerline{\textit{III. All BPS BH in AdS with
horizon of topology $S^3$ are characterized by the same}}
{\textit{\hspace{1cm}constraint among its chemical potentials}.}

\vspace{.5cm}
\noindent In particular, to consider minimal gauge supergravity
solutions we have to set all R-charge chemical potentials equal. Hence the above constraint reduces to
\be \phi = {2\over3}w_+ \quad\hbox{where}\quad \phi_I=\phi
\quad\forall \,I\,. \label{mgc}\ee

\vspace{.5cm}
\textbf{The $S^1\times S^2$ case}: Here we choose to parametrize the NHG again in terms the three $\mu_I$ and a new parameter $\rho$ corresponding to the radius of the $S^1$ with co-ordinate $\phi$. We will see that there is redundancy in this parametrization and only three independent parameters are really need it. Then, we write the NHG in such a way that it has explicitly the $AdS_2\times S^2\times S^1$ factors obtaining,
\bea &&ds^2=v_1\left(-r^2dt^2+{dr^2\over r^2}\right)+v_2\left[d\theta^2+sin^2(\theta)d\psi^2+v_3(d\phi+b)^2\right]\nonumber \\
&&b=-\alpha r dt+b_r{dr\over r}\,,\quad A^I=e^I\cos(\theta)\,d\psi\,,\quad X_I=u_I\,,
\label{brg}
\eea
where $(\psi,\phi)$ have canonical period of $2\pi$ and all the constant components are given in terms of $\mu_I$ and $\rho$ as follows,
\bea
&&u_I={\mu_I\over \gamma_3^{1/3}}\,,\quad v_1={\gamma_3^{2/3}\over \gamma_1^2}\,,\quad v_2={\gamma_3^{2/3}\over (\gamma_1^2-4\gamma_2)}\,,\quad v_3={(\gamma_1^2-4\gamma_2)\over \gamma_3^{2/3}}\rho^2\,,\nonumber\\
&&\alpha={\gamma_3^{1/3}\over\gamma_1\rho}\,,\quad b_r=-{\gamma_3^{1/3}\over\gamma_1\rho}\,,\quad e^I={\mu^I(2\mu^I-\gamma_1)\over (\gamma_1^2-4\gamma_2)}\,.
\eea
As before it is sufficient to reduce on only one $S^1$ to read the form of all the chemical potentials, obtaining the following 4D $U(1)$ gauge fields,
\bea
b=-\alpha r dt+b_r{dr\over r}\,,\quad ^*a_I=-e_I\,r dt\,.
\eea
where the reduced R-charged 4D fields $a_I$ have been Poincare dualized to obtain its electrical version $^*a_I$.

The R-charge chemical potentials are identified with the electric fields of the three $U(1)$ gauge fields that, since we had to take the dual, are related to three dipole charges $q_I$. On the other hand, the identification of the potentials conjugated to the two asymptotic $U(1)$ rotational symmetries is more complicated. In principle, the rotational symmetries of the NHG could be realized as any combination of the two asymptotic $U(1)$ rotational symmetries. Hence, the corresponding conjugated chemical potential we calculate in the NHG could be a combination of the corresponding chemical potential at infinity. To identify this combination, we use the flat supersymmetric Br solution of \cite{Elvang:2004rt} as a guide.

In flat space the Br NHG, can be recast in a form alike (\ref{brg}), by considering the following combination of the two asymptotic $U(1)$ rotational symmetries,
\be
\zeta_2=\phi\qquad\hbox{and}\qquad\zeta_1-\zeta_2=\psi\,,
\ee
From these expressions, we can read the corresponding form of the different chemical potentials to be,
\bea
\begin{array}{ccc}
    \left( \begin{array}{cc}
            q_I \\  J_1 \\ J_2
            \end{array}
    \right) &\longleftrightarrow&  \left( \begin{array}{cc}
            \phi_I=e_I \\  w_1=1-\alpha\\ w_2= \alpha
            \end{array}
    \right)\,.
\end{array}
\eea
where we have used the two constraint that characterized this solution,
\be \sum_I \phi_I =1\qquad\hbox{and}\qquad\ \sum_I \phi_I= 2w_+
\label{brc}
\ee
and as before $w_+=(w_1+w_2)/2$.

We have therefore arrived to the conclusion that the candidate Br solution, if exist has to be constraint such that obeys the above equations. Notice that in flat space the difference between the two angular momenta is proportional to the squared radius of the ring. Here, the difference is constraint to be 1 in units of the $AdS$ radius, and hence is not a free parameter. Also this seems to tell us that there is no un-gauge limit of this Br configuration\footnote{In \cite{NH3} after it was calculated the dipole charge of the NHG, it was found that obey a related constraint that also obstructs the un-gauge limit. Here, we have arrive to the same conclusion using an independent derivation based on the chemical potentials of the putative whole solution.}. In any case, if there is a BPS Br solution in AdS
it has to be a solution of only three independent parameters i.e.

\vspace{.5cm} \centerline{\textit{IV. All BPS Br in AdS with
horizon of topology $S^1\times S^2$ are characterized by the same}}
{\textit{\hspace{.5cm}constraints of (\ref{brc}) among its chemical potentials}.}\vspace{.5cm}

We have said nothing on the possibility of having Bh with only one rotational isometry. The main reason being that is difficult to analyze this option using the Entropy function of Sen, since we know almost nothing of the corresponding NHG. In the next section we explore this possibility using the extension of Bh thermodynamics to extremal cases developed in \cite{yo1,yo2} together with information on the general form of the uplift to 10D.

\section{Thermodynamics \& 10D uplift}
\label{uplift}

As we pointed out before, it is possible to extend the thermodynamics of non-extremal Bh to extremal Bh, as a limiting case at zero temperature. In \cite{yo1} we defined the thermodynamics and the Euclidean action by taking supersymmetric limit of non-extremal Bh solutions, to later in \cite{yo2} show its relation to the attractor mechanism and the entropy function of Sen.

Bh thermodynamics relates space-time concepts with standard thermodynamical variables and laws. As it is well known, the usual non-extremal chemical potentials are related to periodicity conditions in time and in the compact directions supporting angular momentum, where this thermodynamic framework is well behaved if we define all these quantities with respect to co-ordinates not rotating at infinity. For example for $AdS_5$ R-charged rotating Bh, the finite temperature $T$ is related to the inverse radius of the compactified time direction and the chemical potential conjugated to angular momenta are identified with the angular velocities $(\Omega_1,\Omega_2)$ of the rotating horizon. Also, the chemical potentials conjugated to the R-charged gauge fields $\Phi^I$ are related to Wilson loops at the boundary of the solution, but can also be recast as angular velocities along the principal directions $\xi^I$ of the $S^5$ in the full 10D metric. For our studies, it is convenient to uplift the solution such that the different charges and chemical potentials are all handled in the same way \footnote{There are other charges and chemical potentials that can enter in the thermodynamics of Bh, like dipole momenta for Br.} obtaining,
\be t\sim t+{1\over T}\,,\quad \zeta_i\sim\zeta_i+{1\over T}\,\Omega_i\,,\quad \xi_I\sim \xi_I+{1\over T}\,\Phi_I\,.
\ee
In the extremal regime, the different thermodynamic observables are obtained as a limiting case of usual thermodynamic relations, where a careful expansion in the $T\rightarrow 0$ limit should be taken. In particular the extremal chemical potentials ($w_i,\phi_I$) are obtained as the next-to-leading term in the expansion of $T$, as follows
\be
\Omega_i=\Omega^0_i+T w_i+\ldots\,;\quad \Phi_I=\Phi^0_I+T\phi_I+\ldots
\ee
where the leading order corresponds to the extremal value of the horizon angular velocity (that for example in Bh with horizon of spherical topology in $AdS$ is the velocity of light).
After the extremal limit is taken, the periodicity conditions have to be regularized, leaving us with finite new periodic conditions i.e. the new extremal chemical potentials\footnote{Basically, the period of the identifications goes to infinity but there is a finite contribution that remains. The infinity part can be re-absorbed by a co-ordinate transformation in time (see for e.g. \cite{hawking}).}. In our example of above we get
%%%%
\bea \zeta_i\sim \zeta_i+w_i\qquad \xi_I=\xi_I+\phi_I\,. \label{pc1}
\eea
%%%%
On the other hand, this procedure does not defines the chemical potential conjugated to the Energy of our BPS Bh. In principle, the time-like circle opens up and there is no regularized left-over that defines the extremal periodicity. This is related to the normalization problems on the supersymmetric killing vector associated with the time co-ordinate. Fortunately, this is bypassed using the BPS bound that relates the energy with other conserved charges, for example the BPS bound in $AdS_5$ is given by the expression
\be E=J^1+J^2+Q^1+Q^2+Q^3\,.\ee
Then, based in the Euclidean action $I$ (\ref{qsr}), we can deduce the relation between the conjugated potential of $E$ and all other conjugated potentials using basic thermodynamics relations. For example, take the Euclidean action $I$, and exchange some of the charges to include explicitly the energy $E$ and then just read the conjugated potential from the resulting equation i.e.
%%%%
\bea \label{newpotentials}I=w_+ E +
w_-J^-+\lambda_IQ^I-S_{Bh}\,,
\eea
where $w_\pm={1\over2}(w_1 \pm w_2)$, $\lambda_I=(\phi_I-w_+)$ and $J^-=(J_1-J_2)$ and we have chosen to use $(w_+,w_-,\lambda_ I)$ as our set of five independent chemical potentials conjugated to $(E,J_-,Q_I)$ respectively. Hence, we see that in this specific choice the periodicity of $t$ is equal to the periodicity of $\varphi=(\zeta_1+\zeta_2)$ i.e,
%%%%
\bea \label{pc2} t\sim t+w_+.
\eea
%%%%
Looking at the above expression and equation (\ref{pc1}), it is easy to realized that there will be global constraints, if we want to have a well defined Bh thermodynamics. In other words, once we turn on the different chemical potentials of the Bh geometry, we have to check for consistency on the periodic conditions (\ref{pc1},\ref{pc2}) in the solution.

\vspace{.5cm}
\noindent {\bf  1/2 BPS solutions in $AdS_5$} \vspace{.5cm}

Let us first consider minimal gauge supergravity. In \cite{g} Gauntlett et al. calculated the explicit form of the killing spinor of all the solutions that breaks $1/2$ minimal 5D supergravity, that in particular includes the AdS Bh solution. They worked in co-ordinates adapted to the killing vector constructed as a bilinear of the corresponding killing spinor and therefore after uplifted to 10D, the solutions are rotating in all the above named angular directions with velocity of light at $AdS$ infinity.

The killing spinor that characterized all these solutions can be written as follows,
%%%%
\bea \label{spinor} \epsilon=f(r)e^{-{i\over 2}
(\xi_1+\xi_2+\xi_3)}\epsilon_0\,,\eea
%%%%
where $\epsilon_0$ is a constant complex Majorana-Weyl 10d spinor
that satisfies a series of projections not important here and
$f(r)$ is a given function of the radial co-ordinate in $AdS_5$.
Actually, the killing spinor transforms under R-symmetry rotations
as follows
%%%%
\bea \epsilon&=&\epsilon_1+i\epsilon_2 \\\nonumber
\epsilon_1&=&cos(\alpha)\epsilon_1+sin(\alpha)\epsilon_2\\\nonumber
\epsilon_2&=&cos(\alpha)\epsilon_2-sin(\alpha)\epsilon_1 \eea
%%%%
where $\alpha=2\sum_I \xi_I$. Notice that we have to circulate
twice in $\xi$ to arrive to the initial configuration, due to the
$1/2$ factor in (\ref{spinor}). Importantly, due to the fact that our coordinate system is rotating at infinity we velocity of light, we have there that
%%%%
\be \xi_I=t\,. \ee
%%%

To have a well defined thermodynamics, the spinor in the above family of solutions need to be periodically identified i.e. we need to recover the same configuration after we circulate twice on the time loop. But since $\xi_I$ have their own periodic conditions, we get this identification up to its period i.e.
%%%%
\be \sum_I \phi_I =2w_+\,. \label{gc}\ee
%%%%
Notice that, the above analysis is done within minimal gauge supergravity and hence we are constraint to consider all $\phi_I$ equal i.e. $\phi_I=\phi$ and hence we recover the constraint of (\ref{mgc}). Nevertheless, the extension to the $U(1)^3$ model is trivial and is well know how uplift the spinors to 10D, obtaining the same form as before but now with the possibility of consider independent chemical potentials $\phi_I$ recovering the more general case of (\ref{cc}) \footnote{These results seems to be particular to gauge supergravity theories where we have non-trivial gauge transformation for the killing spinors and the BPS bound includes the angular momenta.}.

Summarizing, we have found that imposing consistent boundary conditions on the Killing spinor, once the chemical potentials are introduce, requires an extra constraint among the different chemical potentials. Notice that this extra constraint is above the BPS constraint and is exactly the same constraint we found in the previous section for the NHG with isometry $R\times U(1)\times U(1)$. This is a consistency check on our result, since both derivations are independent in nature. Also, in this section we only used the general form of the killing spinor in the 10D setup and never imposed any restriction on the isometry group of the solutions. Hence we have arrived to a condition stronger than in previous sections,

\vspace{.5cm} \centerline{\textit{V. All BPS Bh in $AdS_5$ are characterized by the same constraints of (\ref{gc})}}
{\textit{\hspace{1.2cm}among its chemical potentials}.}\vspace{.5cm}

We would like to point out that this extra constraint eliminates all the causal problems outside the horizon for the known Bh solutions. Hence, it seems that we can understand our result from a new perspective which tells us that, a well defined supersymmetric Euclidean solution with well behaved thermodynamic properties is incompatible with closed time-like curves \footnote{The fact that there are CTC behind the horizons is irrelevant for the Euclidean regime since the metric terminates smoothly precisely at the horizon.}.

\section{Discussion}
\label{end}
In this work we have derived new general constraint for supersymmetric Bh configurations in the $U(1)^3$-model of 5D gauge supergravity. This constraints come from two different methods;
\begin{itemize}
\item The first one is related to the Entropy function on the NHG, or more precisely to the definition in the NHG of the BPS chemical potentials of the full Bh geometry. The fact that the chemical potentials can be entirely recover with only the NH data is a reminiscence of the Attractor mechanism \footnote{In \cite{yo2}, we show how the different chemical potentials of the full geometry are related to the electric fields in the NHG.} and is the key ingredient in the derivation.
\item The second, is based on the extension of Euclidean methods and in general of Bh thermodynamics to extremal Bh. This extension define extensive properties that translate into global constraints on the Killing spinor.
\end{itemize}
The constraints obtained from the first method are as general as the NHG where the calculations are perform, and therefor only apply to configurations with isometry group $R\times U(1)\times U(1)$. On the other hand, the constraints obtained in the second method are general and should apply to any Bh in the theory. The main result are summarized as follows,
\begin{enumerate}
\item There are no more general BPS Bh with horizon of topology $S^3$ than the ones we already known in \cite{sbh3} for the $U(1)^3$-model.
\item If there is a Br, it has only 3 independent parameters, and the chemical potentials conjugated to the dipole charges are such that $\sum_I \phi^I=1$.
\item All Bh in the $U(1)^3$-model satisfy the constraint $\sum_I \phi^i=2w_+$.
\end{enumerate}
The first result comes from the constraints $I,III$ and $V$. The second, from $I,IV$ and $V$, while the third is $V$.

Let us next consider some speculations regarding the existence of the Br in $AdS_5$. We have found that if the solution exist, it is highly constraint, where the sum of all dipole chemical potentials is equal to one. For Br in flat space this would mean that the radius of the ring is constraint to be equal to the $AdS$ radius. This property is not shared by the spherical Bh cousins although is true that for those the maximal radius is of order of the $AdS$ radius. On the other hand, this constraint on dipole chemical potential seems to define an obstruction to the un-gage limit, since will imply that the sum is actually unbound. A possible solution to this problem is that the Br configuration is asymptotically locally $AdS$, and therefore does not admit the un-gage limit.

We would like to add some comments on the possible extension of the above methods to more general settings like the 5D $SO(6)$ gauge supergravity and the full type IIB supergravity with $AdS_5\otimes S^5$ fixed asymptotic conditions. In both cases the extended thermodynamic properties should provide global constraints in the corresponding killing spinor of putative Bh solutions. In particular for the $SO(6)$-model, gauge transformations are non-abelian any more but still are related to $SO(6)$ isometries in the uplift configuration, and contain the shift in the 3 principal directions $\xi^I$ as the Cartan abelian subalgebra. Hence, it looks like the constraint of (V) will be present in this setting too. In the type IIB case, since the solutions are still bound to be asymptotically $AdS_5\otimes S^5$, we have that all the solutions have to be labeled by the same chemical potentials that we have discussed along this article. Also, all the killing spinor should have the same asymptotic behavior so it seems possible to define the corresponding constraints based only on the asymptotic general form.

In the dual theory, the picture is far from clear. In principle Bh should be related to supersymmetric ensembles labeled by the corresponding chemical potentials. In \cite{index} it was calculated an Index, that turns out to depend on only 4 chemical potentials. Unfortunately this Index is blind to Bh configurations due to strong cancellations between bosonic and fermionic degrees of freedom, so is not of much help. Also, in this same article, it was calculated the explicit form of the free partition function for $1/16$ BPS ensembles, but again was difficult to compare with the Bh solution due to mainly two facts; first because it is a free theory calculation and second because the resulting partition function is not constraint and hence depends in all 5 chemical potential. In \cite{yo1}, we shed light on this issue since we found that the extra Bh constrain of (\ref{gc}) is exactly the same constraint that characterized the Index. We use it on the free partition function obtaining strong similarities with the Bh including the same phase diagrams and first order phase transitions.

At this point, it is clear that the constraint appearing in the definition of the Index in the dual SYM is relevant to Bh physics since it reproduces the behavior of the Bh partition function and we have seen in this article that it is a common feature of all Bh, at least in the $U(1)^3$-model. In fact putting together the information coming from supergravity side and the SYM side, we propose the natural conjecture,

\vspace{.5cm} \centerline{\textit{All Bh in type IIB supergravity with $AdS_5\otimes S^5$ fixed asymptotic conditions}}{\textit{\hspace{.7cm} should satisfy the constrain (V).}}\vspace{.5cm}

As a last point, we would like to add that these methods could be applied to other supergravity theories, not necessarily gauged. Certainly the attractor mechanism is a common feature of all these theories, and as for example we showed in \cite{yo2}, there is no problem in the corresponding extension of the thermodynamics properties for the extremal Bh. For example, in $U(1)^3$-model of 5D supergravity we have basically only two cases, the BPS BMPV Bh that has only one angular momentum and the generalization to the case of two angular momenta corresponding to the Br. Here, the Br is label by 5 independent parameters and hence there is no common overall constraint among the 5 conserved charges. In this case, the NHG analysis will produce the constraints that are particular to each different NHG topology.

%%%%%%%%%%%%%%%%%%%%%%%%%%%%%%%%
\vspace{1.5cm} \noindent
{\bf Acknowledgements}\\

{\small We would like to thanks the organizers of the pre-string 2007 meeting in Granada where part of this work was developed. Also, we thanks  Marco Caldarelli, Alesio Celi, Roberto Emparan, Juan Maldacena, J. Morales and in particular to O.~J.~C.~Dias for their clarifications and key discussions to. Finally we thanks Julia and Helena for critical reading of this manuscript.\vspace{.3cm}

This work was partially funded by the Ministerio de
Educacion y Ciencia under grant CICYT-FEDER-FPA2005-02211 and CSIC under the I3P program.}

%%%%%%%%%%%%%%%%%%%%%%%%%%%%%%%%

%%%%%%%%%%%%%%%%%%%%%%%%%%%%%%%%
\end{document}